\documentclass[12pt,preprint]{aastex}
\usepackage{color}
\shorttitle{Neutrino and CR Release from GRBs}
\shortauthors{Asano \& M\'esz\'aros}

\begin{document}

\title{
Neutrino and Cosmic-Ray Release from Gamma-Ray Bursts:
Time-Dependent Simulations
}
\author{\scshape Katsuaki Asano\altaffilmark{1},
and
Peter M\'esz\'aros\altaffilmark{2}}
\email{asanok@icrr.u-tokyo.ac.jp, nnp@psu.edu}

\altaffiltext{1}{Institute for Cosmic Ray Research, The University of Tokyo,
5-1-5 Kashiwanoha, Kashiwa, Chiba 277-8582, Japan}
\altaffiltext{2}{Department of Astronomy \& Astrophysics; Department of Physics;
Center for Particle \& Gravitational Astrophysics;
Pennsylvania State University, University Park, PA 16802}

\date{Submitted; accepted}

\begin{abstract}

We revisit the neutrino and ultra high-energy cosmic ray (UHECR) production
from gamma-ray bursts (GRBs)
with time-dependent simulations for the proton-induced cascades.
This method can generate self-consistent photon, neutrino and 
escaped neutron spectra.
To obtain the integrated background spectra, we take into account the 
distributions of the burst luminosity and pulse duration timescale.
A benchmark case with standard GRB luminosity function, a bulk Lorentz 
factor $\Gamma=300$ and a proton to gamma-ray luminosity fraction 
$f_{\rm p}=10$, is consistent with both the neutrino upper-limits and 
the observed UHECR intensity at $\sim 10^{20}$ eV,
while requiring a different type of UHECR source at the ankle.
For the benchmark case the 
GRBs in the bright end of the luminosity function, which contribute 
most of the neutrinos, have their photon spectrum substantially distorted
by secondary photons. Such bright GRBs are few in number, and reducing 
their $f_p$ eliminates the distortion, while reducing the neutrino production. 
Even if we neglect the contribution of the brightest GRBs, 
the UHECR production rate at GZK energies is almost unchanged. 
These nominal GRB models, especially with $L_{\rm iso} \lesssim 10^{53} 
~\mbox{erg} ~\mbox{s}^{-1}$, appear to meet the current constraints as 
far as being candidate UHECR sources above the ankle energy.

\end{abstract}


\maketitle

\section{Introduction}
\label{sec:intro}

Since the pioneering study by \citet{wax97},
the possibility of neutrino emission from gamma-ray bursts (GRBs)
has been discussed by many authors \citep[e.g.][and references therein]{mur06,hum12}
in the context of ultra high-energy cosmic ray (UHECR) source
\citep{wax95,vie95}.
The IceCube team provided upper-limits for the neutrino intensity
from GRBs \citep{abb12}, which is still above the prediction
in some fiducial models \citep{he12,li12,hum12}.
While the extra spectral components in the GeV band detected
with {\it Fermi} \citep{510,902B} may indicate
possible signatures of photopion production by accelerated protons
\citep{asa09a,asa10},
the non-detection of neutrinos from the very bright burst
GRB 130427A constrains the proton fraction to the gamma-ray luminosity
\citep{gao13}.
Besides models based on the classical internal shock paradigm,
the neutrino emission from alternative models such as
dissipative photospheres \citep{gao12}
or ICMART \citep{zha13} has also been discussed.
In addition to these models, neutrinos from low-luminosity
GRBs \citep{mur06}, or ultra-long GRBs \citep{mur13} etc., 
are also interesting as potential sources for the detection
of cosmological PeV neutrinos with IceCube \citep{aar13}.

However, the internal shock for GRBs  is still the archetypal 
and most widely considered model for producing UHECRs.
In this paper, we revisit the neutrino and UHECR production
in the standard internal shock model, using our time-dependent
method of \citet{asa12} \citep[see also][]{asa11}
to simulate the proton-induced cascade process.
The advantages of our time-dependent code are
1) consistent spectra of photons and neutrinos,
and 2) a more realistic treatment for the neutron escape.
As the number of secondary photons increase, the pion production
efficiency is enhanced. This non-linear process affects the amount 
of neutrons and the spectrum.
Especially in cases where the protons experience
multiple collisions with photons, strong cooling
due to pion production occurs before neutrons escape, which
leads to suppression of the UHECR amount.
Although our code is based on a one-zone approximation,
it allows for the time-dependent change of the size, densities 
and other variables, and the effect of this on the gradual escape 
of neutral particles can be simulated.

In this paper, we take into account the distributions of the luminosity 
and the pulse width or variability timescale to obtain the neutrino and 
UHECR spectra, which are propagated from cosmological distances with a
Monte Carlo approach. The exact UHECR escape mechanism is unknown, and this 
can affect significantly the resultant spectra of neutrinos and UHECR 
\citep[see e.g.][]{bae14}.  Two extreme models for the UHECR escape, 
a pessimistic one and an optimistic one, are considered here.

\section{Simulations}
\label{sec:model}

\begin{figure}[htb!]
\centering
\epsscale{1.0}
\plotone{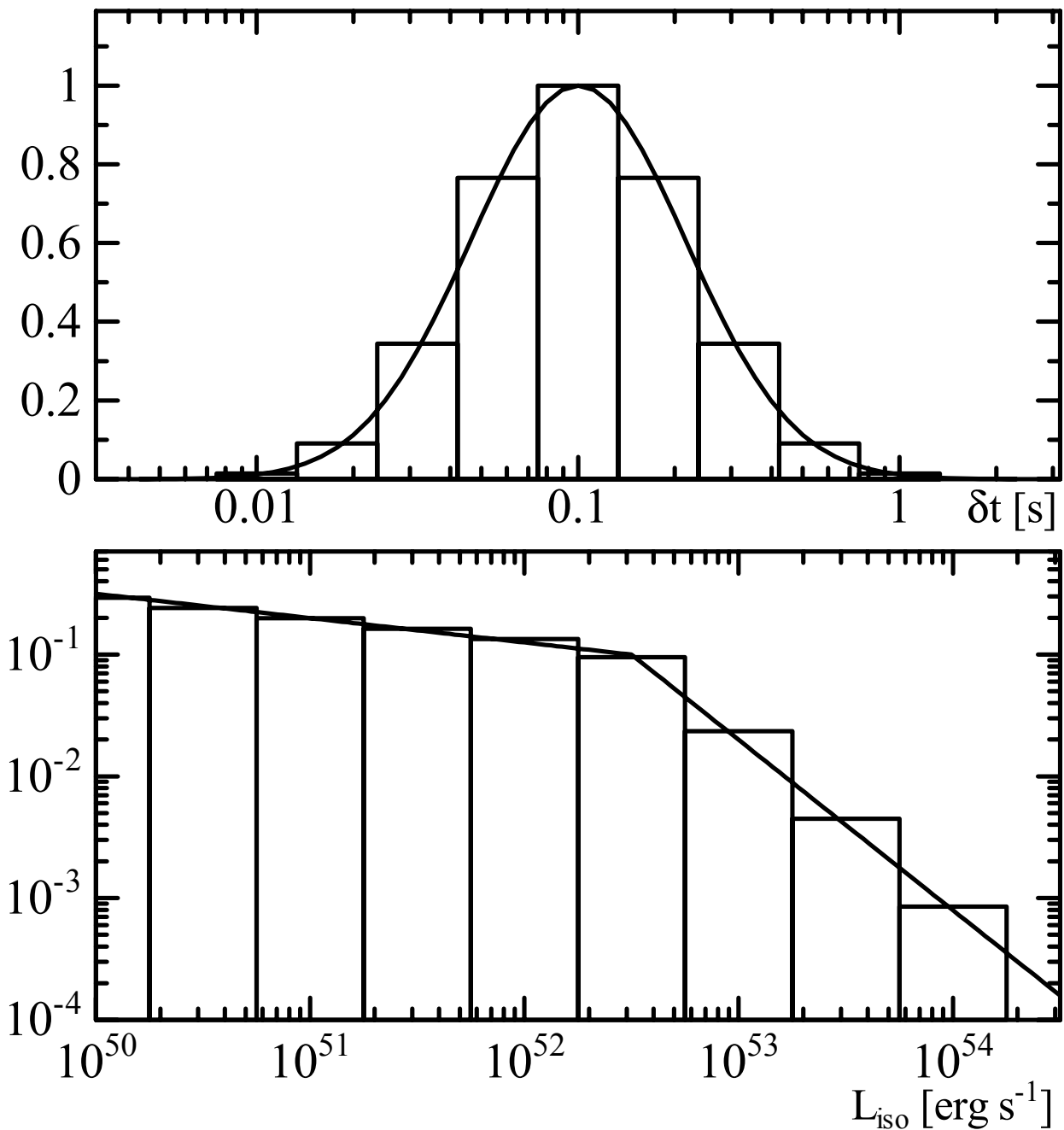}
\caption{The assumed variability (upper) and luminosity (lower) distributions (see text).
\label{fig:1}}
\end{figure}

The details of our numerical code to simulate the photon and neutrino emission
is discussed in \citet{asa12}.
Here, as a benchmark case, we fix the bulk Lorentz factor at
$\Gamma=300$. We consider GRBs with isotropic-equivalent
luminosities $L_{\rm iso}$ greater than $10^{50}~\mbox{erg}~\mbox{s}^{-1}$.
The luminosity function per logarithmic interval is taken from \citet{wan10}, as
$\phi(L) \propto L^{-0.17}$ below $L_{\rm iso}=10^{52.5}~\mbox{erg}~\mbox{s}^{-1}$,
and $\phi(L) \propto L^{-1.44}$ above that.
Following this function, we divide the luminosity into 9 intervals,
as shown in Figure \ref{fig:1}.
Another important parameter for the neutrino emission
is the initial shock radius $R_0$ at which the proton injection starts.
This radius $R_0$ is related to the variability timescale 
as $\delta t=R_0/(2 c \Gamma^2)$.
Our time-dependent code provides as output the lightcurve for one shell
with the Doppler and curvature effects.
We changed $R_0$ from $10^{14}$ to $10^{16}$ cm,
which corresponds to $R_0/(2 c \Gamma^2)=0.019$--$1.9$ s.
The obtained lightcurves show a sharp rise and long tail
\citep{asa11}, the rise timescale being slightly
shorter than the above simple estimate.
According to our results,
the radii $R_0$ are allocated to 9 timescale bins as shown in Figure \ref{fig:1}.
\citet{nak02} showed that the pulse width follows
a log-normal distribution with the parameters $\mu=0.065$ ($\delta t \simeq 1$ s)
and $\sigma=0.77$.
The pulse width is wider than the rise timescale we need.
In the study of \citet{bha13}, the minimum variability timescale is
typically $0.25$ s. Therefore, taking into account cosmological redshift effect,
we shift the distribution peak in \citet{nak02} to $\delta t = 0.1$ s
with the same $\sigma$ to obtain the rise timescale distribution.
In Figure \ref{fig:1}, the histogram for the $\delta t$-distribution assumed
is shown with a log-normal distribution.

Given $L_{\rm iso}$ and $R_0$,
the initial photon energy density in the shell frame
is written as $ L_{\rm iso}/4 \pi c R_0^2 \Gamma^2$.
The magnetic energy density is assumed to be $10$ \% of the photon energy density.
In this paper, we omit to simulate the primary photon production,
because the emission mechanism is not well understood, there being several 
competing models.
The photon spectrum is simply assumed to be the conventional Band function, 
whose spectral peak energy $\varepsilon_{\rm p}$ satisfies the 
$\varepsilon_{\rm p}$--$L_{\rm iso}$ relation \citep{yon04}.
We express this as $\varepsilon_{\rm p}=10^{-22.97} L_{\rm iso}^{0.49}$ keV
following \citet{nav12}, where $L_{\rm iso}$ is in cgs unit.
The photon indices are fixed as $\alpha=-1.0$ and $\beta=-2.25$.
The shell width in the comoving frame is taken as $W'=R_0/\Gamma$,
which gives us the total photon energy from one pulse $E_{\rm ph}$.
To satisfy the typical total energy of the burst $E_{\rm iso}$,
we need multiple pulses for one burst.
We estimate the average pulse numbers using a $E_{\rm iso}$--$L_{\rm iso}$
based on the sample in \citet{ghi12} (see below).
We inject protons of total energy $f_{\rm p} E_{\rm ph}$ 
in a timescale $W'/c$ at a constant rate\footnote{Other authors define proton energy
in terms of electron energy, $f_{\rm p} E_{\rm e}$, which is equivalent to our definition
for fast cooling electrons.}.
Hereafter, we adopt $f_{\rm p}=10$ as a benchmark case.
The proton number spectrum at injection is assumed to be $\propto \varepsilon^{-2}
\exp(-\varepsilon/\varepsilon_{\rm max})$, where
the maximum proton energy $\varepsilon_{\rm max}$
is calculated with the Bohm limit assumption,
taking into account the cooling due to synchrotron and photomeson production.
Our time-dependent code follows the cascade processes in the shell
as far as $R=30 R_0$. The primary photons and the secondary
photons/neutrinos gradually escape from the shell.

\begin{figure}[htb!]
\centering
\epsscale{1.0}
\plotone{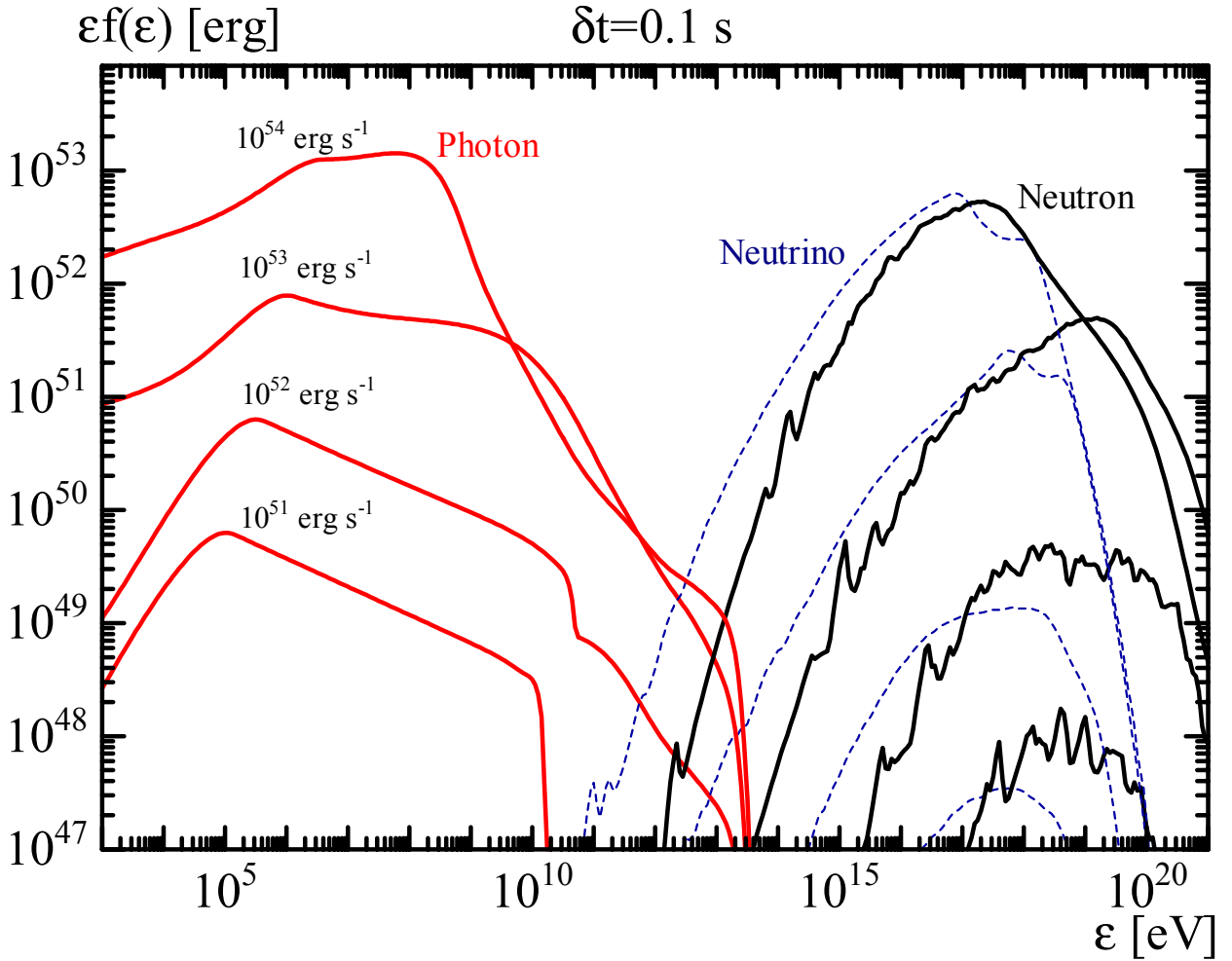}
\caption{Photon (solid red), neutron (solid black),
and neutrino (dashed blue; $\nu_\mu$,${\bar\nu}_\mu$,$\nu_e$,${\bar\nu}_e$) 
spectra escaping from one shell
for the neutron conversion model, for one particular variability 
timescale of $0.1$ s and for various luminosities. 
\label{fig:2}}
\end{figure}

In total we carried out 81 runs changing $R_0$ and $L_{\rm iso}$
to simulate the emission from one shell.
The emission from a burst arises from multiple shells.
Here, we simply multiply the average number of pulses for one burst.
For the cosmic ray release, we consider two extreme cases:
the neutron conversion model and the sudden release model.
The neutron conversion model is the most pessimistic model,
in which only neutrons can escape from the shell.
In the sudden release model, all protons and neutrons in the shell
are released at $R=3 R_0$, which is the most optimistic case for cosmic 
ray production; while  artificial, we consider it as a limiting case.
The realistic cosmic-ray escape may be between those two extreme cases.
Figure \ref{fig:2} shows examples of the time-integrated spectra
of photons, neutrinos and neutrons released from one shell.
The photon spectrum for
$L_{\rm iso}=10^{54}~\mbox{erg}~\mbox{s}^{-1}$ is dominated
by the secondary photons originating from proton cascades.
In this ``proton-dominated'' case \citep{asa09b},
the GeV flux is brighter than the MeV flux,
and the low-energy spectrum is soft and curved,
the photon index changing from $\sim -1.8$ to $-1.6$.
Such photon signatures have not been identified as common
properties for luminous GRBs.
From this viewpoint, we emphasize that the benchmark case of 
$\Gamma=300$ and $f_{\rm p}=10$ seems to be ruled out for bursts 
brighter than $L_{\rm iso}=10^{54}~\mbox{erg}~\mbox{s}^{-1}$.
However, we keep this assumption temporarily,
and re-discuss this problem further below.

The neutrino spectra in Figure \ref{fig:2} show complex changes
with luminosity. This is partially because of the importance
of the proton/neutron cooling
and synchrotron cooling of muons/pions, which grow with luminosity.
Note that the neutrino spectrum is a summation of
both the muon- and pion-decay contributions.

\section{UHECRs and Neutrino Background}
\label{sec:UHECR}

Based on the results in the previous section, we calculate the cumulative 
contributions of GRBs over redshifts of $z \leq 5$ to the UHECR and neutrino fluxes.
We estimate the average number of pulses per burst from a relation
between $E_{\rm iso}$ and the peak $L_{\rm iso}$.
For this purpose, we use the same relation as assumed in \citet{kak12},
\begin{eqnarray}
\log_{10}
\left( \frac{E_{\rm iso}}{10^{52}
\mbox{erg}} \right)=0.56+1.1 \log_{10} \left( \frac{L_{\rm iso}}{10^{52}
\mbox{erg}~\mbox{s}^{-1}} \right),
\end{eqnarray}
which is obtained from the GRB sample in \citet{ghi12}.
The GRB rate per comoving volume is also taken from \citet{wan10},
$R_{\rm GRB}(z) \propto (1+z)^{2.1}$ for $z \leq 3.0$
and $\propto (1+z)^{-1.4}$ for $z>3.0$.
with the local GRB rate of $1.3~\mbox{Gpc}^{-3}~\mbox{yr}^{-1}$
above $10^{50}~\mbox{erg}~\mbox{s}^{-1}$.
\citet{wan10} found no evidence of a luminosity evolution,
so we do not consider evolution here.
For the cosmological parameters we adopted $h=0.7$, $\Omega=0.3$,
and $\Lambda=0.7$.

The probability distributions for $\delta t$ and $L_{\rm iso}$
together with the average pulse number
give us the average neutrino (UHECR) spectrum $N_\nu (\varepsilon)$
($N_{\rm CR} (\varepsilon)$) for one GRB.
For simplicity, we neglect the dispersions in the $E_{\rm iso}$--$L_{\rm iso}$
and Yonetoku relations.
We calculate the spectral intensity of the neutrino (UHECR) background
in the standard way \citep[see e.g.][]{ber06,ahl11}.
For the cosmic ray propagation, we take into account the energy loss
due to the Bethe--Heitler pair production and photopion production.
We adopt the model of \citet{kne04} for the extra galactic background light.
The calculation method is basically the same Monte Carlo method as that for 
the cascade calculation inside the GRB shell.

\begin{figure}[htb!]
\centering
\epsscale{1.0}
\plotone{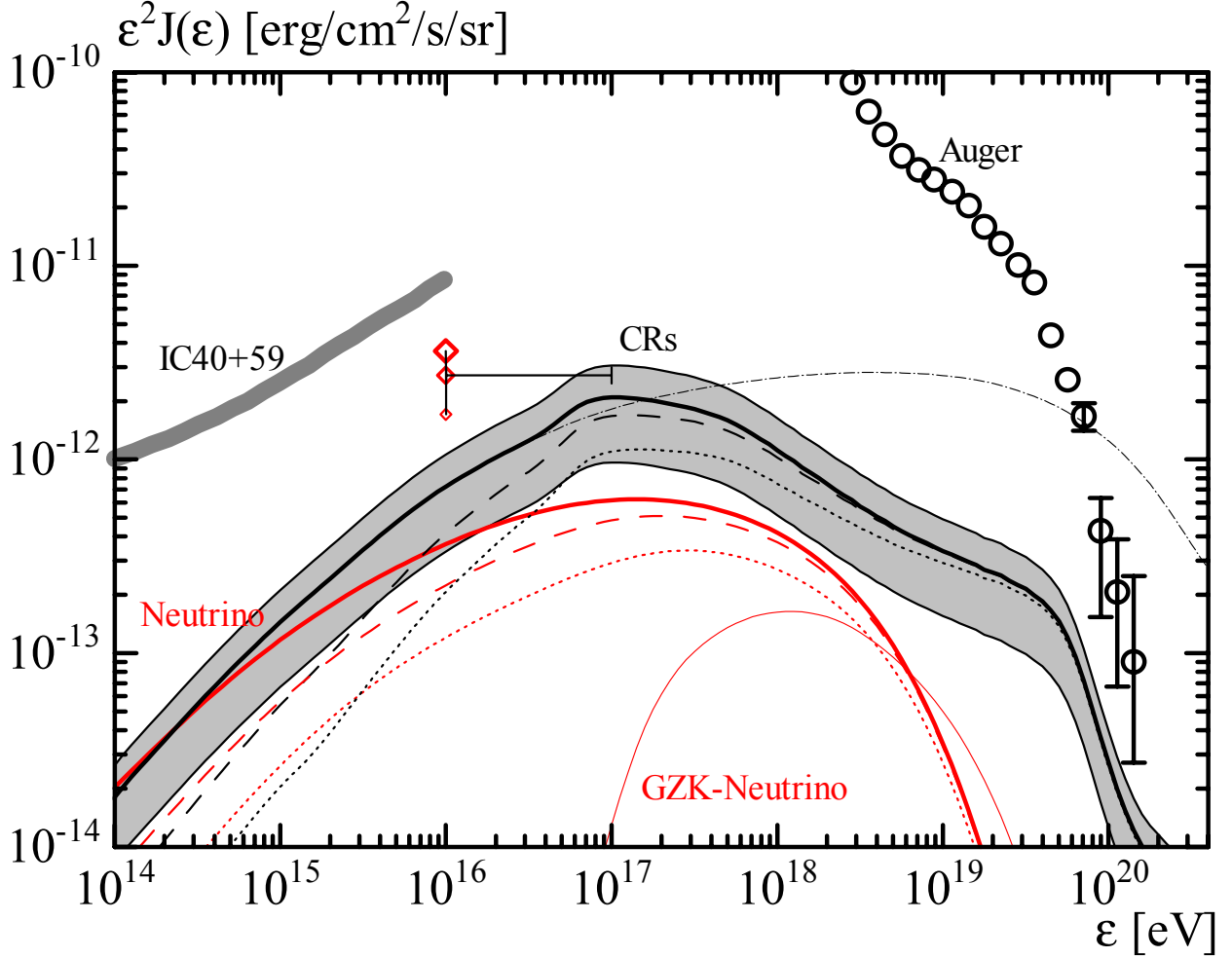}
\caption{The CR (black) and neutrino (red, $\nu_\mu$ and $\bar{\nu_\mu}$
after oscillation) 
diffuse intensities for the neutron conversion model.
The gray shaded area for CRs
indicates the uncertainty in the local GRB rate.
The thin dash-dotted line is the CR spectrum
without the effects of photomeson production
and Bethe--Heitler pair production.
The data for the UHECR intensity (circles) is from \citet{sch13}.
The dashed (dotted) lines are the spectra without the contribution of GRBs
of $L_{\rm iso} \geq 10^{54}$ ($\geq 10^{53.5}$) $\mbox{erg}~\mbox{s}^{-1}$.
The cosmogenic neutrino spectrum produced via the GZK process is also shown
as the thin red line.
The gray thick line is the neutrino upper-limits in \citet{abb12},
which are function of the neutrino break energy,
assuming a spectral shape as $\propto \varepsilon^{-1}$
below and $\propto \varepsilon^{-2}$ above.
While this can be regarded as approximate diffrential upper-limits,
our results have no distinct break in the spectra.
For reference, we also plot the integrated energy fluxes (diamonds)
for the three results with uncertainty in break energies of $10^{16}$--$10^{17}$ eV.
}
\label{fig:3}
\end{figure}

\begin{figure}[htb!]
\centering
\epsscale{1.0}
\plotone{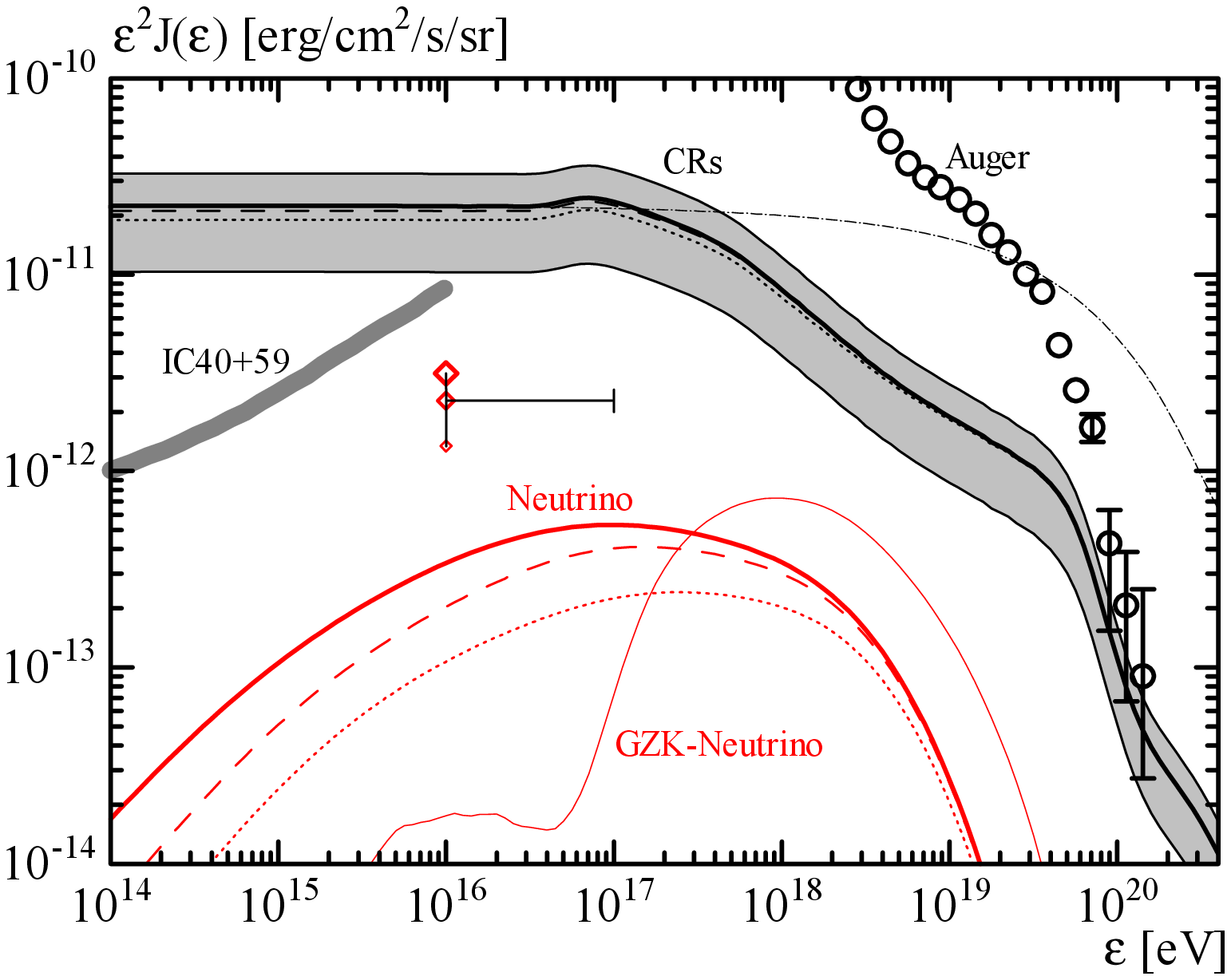}
\caption{Same as Figure \ref{fig:3} but for the sudden release model.
\label{fig:4}}
\end{figure}

Figures \ref{fig:3} and \ref{fig:4} show the resultant spectra
for the neutron conversion model and the sudden release model, respectively.
In both models, the neutrino intensities are well below the upper-limit of 
\citet{abb12},  owing to the relatively long average $\delta t$
and the inclusion of time-dependent effects.
The results are syntheses of the neutrino emission
for various luminosities, variabilities, and redshifts.
Therefore, the spectral shapes are curved, and have no clear break,
but are roughly consistent with the results in \citet{he12}.
Although the CR spectral shapes are significantly different in the neutron escape
and sudden release models, the intensities around the highest energy range 
($\sim 10^{20}$ eV) are similar.
Even for the conservative assumptions in the benchmark models,
GRBs appear able to contribute significantly to the higher energy CR flux.
However, the intensities at the ankle region ($\sim 10^{18.5}$ eV)
are far below the observed ones, requiring different sources.

In Figures \ref{fig:3} and \ref{fig:4} we also plot the diffuse spectra
neglecting the contribution of GRBs in the highest luminosity-bin
of Figure \ref{fig:1} ($\sim 10^{54}~\mbox{erg}~\mbox{s}^{-1}$).
As discussed in the previous section, the photon spectrum
for such luminous GRBs is inconsistent with observations.
At least for these luminous GRBs, consistency with the photon 
(and UHECR as well as neutrino) observations might be achieved
if the pion production is somehow reduced, e.g. 
by decreasing $f_{\rm p}$ or increasing $\Gamma$.
This would also agree with the non-detection of neutrinos from
the very bright burst, GRB 130427A \citep{gao13}.
The number fraction of bursts in the highest luminosity-bin is only $0.085$ \%.
Nevertheless, ignoring this bin the neutrino intensity in the low-energy region
is significantly suppressed (see the dashed lines),
while the highest energy CR spectrum is almost unchanged.
If we additionally ignore the contribution of one more luminosity-bin
at $\sim 10^{53.5}~\mbox{erg}~\mbox{s}^{-1}$ ($0.45$ \% in the number fraction),
the neutrino background in the energy range constrained by IceCube 
is even further suppressed (dotted lines).
In other words, current neutrino limits may have constrained the emission from 
only a small fraction of bright GRBs, as also suggested by \citet{he12}
and \citet{adrian13} (e.g. their Fig. 3b).

As is well known, the neutrino production efficiency
is sensitive to the bulk Lorentz factor $\Gamma$.
If the average of $\Gamma$ is an increasing function of $L_{\rm iso}$
or $E_{\rm iso}$, as discussed in \citet{he12}, the
neutrino production would be suppressed even for a higher $L_{\rm iso}$.
This is consistent with the subtraction of the most luminous GRBs
that we discussed above.
If the less luminous GRBs ($L_{\rm iso} \leq 10^{53}~\mbox{erg}~\mbox{s}^{-1}$)
have properties similar to those of the benchmark model,
they can contribute significantly as UHECR sources at $\sim 10^{20}$ eV
as shown in the Figures \ref{fig:3} and \ref{fig:4}.
Such relatively low-luminosity GRBs may in principle have larger $f_{\rm p}$
without infringing the same energy budget constraints which affect the brightest GRBs. 
Thus, in principle, if $f_{\rm p} \gtrsim 100$ for bursts of $L_{\rm iso} \leq 
10^{53}~\mbox{erg}~\mbox{s}^{-1}$ in the sudden release model, such GRBs could 
be a dominant UHECR sources above the ankle energy \citep{bae14}.


\section{Discussion}
\label{sec:sum}

We have considered a generic internal shock model of GRBs with an observationally
motivated distribution of luminosities and variability timescales. The benchmark 
case  with Lorentz factor $\Gamma=300$ and proton load $f_{\rm p}=10$ is consistent
with both the neutrino upper-limits by IceCube and the observed UHECR intensity at 
$\sim 10^{20}$ eV.  However, the photon spectrum for very bright GRBs with 
$L_{\rm iso} \sim 10^{54}~\mbox{erg}~\mbox{s}^{-1}$ seems inconsistent with observations.
A lower $f_{\rm p}$ or higher $\Gamma$ would be required for such luminous GRBs,
which would further reduce their neutrino contribution. However, even if we neglect 
the contribution of the brightest GRBs in the benchmark case, the predicted UHECR 
intensity at $\sim 10^{20}$ eV is almost unchanged.
For relatively less luminous GRBs, the luminosity function and
$E_{\rm iso}$--$L_{\rm iso}$ relation may have a large uncertainty.
In terms of the energy budget, a larger proton loading such as $f_{\rm p}=100$
can be allowed for such less luminous GRBs.  Therefore, internal shock models
of GRBs, especially those with $L_{\rm iso} \lesssim 10^{53}~\mbox{erg}~\mbox{s}^{-1}$,
are compatible with the observational constraints so far, and can be UHECR source 
candidates above the ankle energy.

Our calculation of the GRB internal shock takes into account explicitly the 
time-dependent effects of the shock region  expansion. This differs from previous 
GRB internal shock neutrino and UHECR calculations, including e.g.
\citet{bae14}. However, despite differing in method, our results are broadly 
compatible with theirs, once these differences are taken into account.
While we have incorporated explicitly the distributions in luminosity and variability,
they have performed separate calculations for a range of parameters. 
They considered several models for the redshift evolution, while we have taken
the \cite{wan10} distribution as representative. 
Two other differences are that secondary photon production is taken into account 
in our model, which enhances the pion production; 
and our time-dependent code provides an improved neutron escape treatment.
The latter can lead to appreciable differences especially in cases where the 
protons experience multiple collisions with photons, which lead to strong cooling due 
to pion production before neutrons escape. Multiple collisions are treated approximately 
in their scheme, while they are included explicitly in our time dependent scheme.
Our choice of magnetic to photon energy fraction of 10\% versus theirs of unity
and our stronger cooling due to secondary photons leads to a somewhat larger prompt to 
GZK neutrino ratio than theirs.
They considered a neutron escape and a leakage dominated model for CR escape,
while we considered a time-dependent neutron escape and an extreme sudden release model.
In the leakage dominated model, the highest energy CRs escape freely, while lower energy CRs 
also escape according to an escape probability that is proportional to the energy.
In agreement with \citet{bae14} we find that the less luminous bursts are less
constrained. However the inclusion in our calculation of the time-dependence of 
the shocked shells as they expand enhances the neutron and CR escape, reducing 
the neutrino production. Thus, for accelerated protons, in our case the same models
appear less constrained even for standard parameters\footnote{If the UHECR were dominated
by heavy elements, photo-dissociation in the internal shock GRB environment can be
reasonably avoided \citep{Wang+08nuc,Horiuchi+12nucjet}, and the predicted
diffuse neutrino flux is much lower than for protons, e.g. \cite{Anchordoqui+08nuc,
Murase+10cosmonucomp}}.

We note that the diffuse PeV neutrino flux detected by IceCube \citep{aar13} cannot 
be reproduced by the GRB internal shock models discussed here. On the other hand,  these
models are not constrained by the diffuse PeV-EeV $E^{-2}$ or model-independent 
IceCube limits \citep{aar13b}, which are an order of magnitude above
our predicted flux in this energy range.

The present calculations suggest that the current IceCube TeV neutrino limits
constrain mainly the small fraction of the bright end GRBs, as far as their 
potential role as UHECR sources.
However, the observed gamma-ray spectra for such bright GRBs,
in the context of internal shock models of UHECR, suggests that these 
bright bursts must then have an inefficient neutrino production. 
It may be natural to consider that luminous GRBs have larger $\Gamma$,
as suggested by {\it Fermi} observations \citep{916C}.
In this case, the current non-detection of neutrinos would be a natural consequence.

\begin{acknowledgments}
First we appreciate the anonymous referee for valuable advise.
We appreciate comments from M. Ahlers, P. Baerwald and  S. Yoshida
and discussions with K. Murase and H. Takami.
We acknowledge partial support by Grants-in-Aid for Scientific Research
No.25400227, 24540258 from the Ministry of Education, Culture, Sports, 
Science and Technology (MEXT) of Japan, and NASA NNX 13AH50G.
\end{acknowledgments}

\end{document}